\documentclass[11pt]{article}
\usepackage{amssymb,amsmath}

\newcommand{\be}{\begin{equation}}
\newcommand{\ee}{\end{equation}}
\newcommand{\bea}{\begin{eqnarray}}
\newcommand{\eea}{\end{eqnarray}}
\def\1#1{^{(#1)}}

\def\be{\begin{equation}}
\def\ee{\end{equation}}
\def\bea{\begin{eqnarray}}
\def\eea{\end{eqnarray}}
\def\p{\prime}
\def\SCR{\tiny}
\def\DEV[#1]{\mathaccent'27{#1}}
\def\V#1{\bf {#1}} 
\newcommand{\Grad}{{\B \nabla}}
\def\PDX[#1,#2]{ \ifx{#1}{#2} {\partial \over \partial x}
        \else \ifx{#1}\NOTH {\partial {#2} \over \partial x}
         \else  {\partial {#2} \over \partial x_{#1} } \fi \fi}
\def\B#1{{\makebox{\boldmath ${#1}$}}}
\newcommand{\bitem}{\begin{list}{$\bullet$}{\setlength{\itemsep}{-0.4mm}}}
\newcommand{\eitem}{\end{list}}
\begin{document}
\title{Effects of Material Symmetry on the Coefficients of Transport
\\ in Anisotropic Porous Media}
\author{Jacob Bear, Leonid G. Fel and Yoram Zimmels \footnote{Two months
after completing this paper, our colleague Prof. Yoram Zimmels passed away.}
        \vspace{0.4cm}\\
 Department of Civil and Environmental Engineering,\\
 Technion -- Israel Institute of Technology, Haifa, 32000 Israel}
\date{}
\maketitle
\begin{abstract}
The objective of this article is to highlight certain features of a number of
coefficients that appear in models of phenomena of transport in \textit{anisotropic}
porous media, especially the \textit{coefficient of dispersion,} the 2nd rank tensor
$D_{ij}$, and the \textit{dispersivity coefficient}, the 4th rank tensor $a_{ijkl}$,
that appear in models of solute transport. Although we shall focus on the transport
of mass of a dissolved chemical species in a fluid phase that occupies the void
space, or part of it, the same discussion is also applicable to transport
coefficients that appear in models that describe the advective mass flux of a fluid
and the diffusive transport of other extensive quantities, like heat. The case of
coupled processes, e.g. the simultaneous transport of heat and mass of a chemical
species, are also considered. The entire discussion will be at the
\textit{macroscopic level}, at which a porous medium domain is visualized as
homogenized continuum.
\end{abstract}
\section{The coefficient of dispersion, ${\sf D}_{ij}$}
\label{sec:1}

We consider the transport by the mechanism of dispersion of a solute in a fluid
that occupies the void space of a porous medium, or part of it. The coefficient
of dispersion, ${\sf D}_{ij}$, appears in the Fickian-type expression for the
dispersive flux of a solute (e.g. Bear, 1961)
 \be
 J_i=- {\sf D}_{ij} \PDX[j,c],\qquad i,j \equiv x,y,z,
 \quad \mbox{or} \quad i,j \equiv 1,2,3,
 \label{flux-c}
  \ee
  where $J_i$ denotes the $i$th component of the solute flux vector ${\V J}$
($\equiv$ solute mass passing through a unit area of fluid in the porous medium
cross-section, per unit time), and $c$ is the solute's concentration (= mass
of solute per unit volume of fluid). Furthermore, we use the term \textit{flux}
as an abbreviation for \textit{flux density}. In (\ref{flux-c}), and everywhere
else in this article, \textit{Einstein summation convention} is applicable,
unless the sum symbol is used.
The coefficient ${\sf D}_{ij}$ is a 2nd rank tensor, relating the vector ${\V
J}$  to the vector $\Grad c$.

Equation (\ref{flux-c}) is valid for the general case of an anisotropic porous
medium, with the isotropic medium as a special case. The dispersion coefficient
is characterized by:

\begin{enumerate}
\item In thermodynamics, the \textit{rate of entropy production},
$\dot{{\cal S}}$, is related to the thermodynamic driving force, ${\V X}$, and the
thermodynamic flux, ${\V Y}$, called "conjugated flux and force" (De Groot and
Mazur, 1962), by $\dot{{\cal S}}=Y_i X_i$. Here, the flux of the solute, ${\V J}$ is
driven by $-\Grad c$, which acts as a "driving force." In this case, the rate of
entropy production can be expressed by
 \be
 \dot{{ \cal S}} = \chi \left[(-{\sf D}_{ij} \PDX[j,c])\right]
                        \times \chi \left[ (- \PDX[i,c])\right] \geq 0,\quad
                                          \mbox{or} \quad
     {\sf D}_{ij} \PDX[j,c]\PDX[i,c] \geq 0,
    \label{sigma-1}
    \ee
in which, ${\V Y}=\chi {\V J}$ and ${\V X} =-  \chi \Grad c$. The (dimensional)
parameter, $\chi$, depends on the considered transport phenomenon. Hence, the ${\sf
D}_{ij}$--matrix is \textit{non-negative definite}.
\item The ${\sf D}_{ij}$--matrix is \textit{symmetric}, i.e.,
\be
 {\sf D}_{ij} = {\sf D}_{ji}.
   \label{dij-symm}
 \ee
\end{enumerate}
The above two statements are consequences of the fact that  ${\V X}$ and ${\V J}$
are \textit{thermodynamically conjugated force and flux} (De Groot and Mazur, 1962),
i.e., they  satisfy
 \be
   \frac{\partial J_i}{\partial X_j}= \frac{\partial
J_j}{\partial X_i}. \label{xj}
   \ee

 We may comment here that the above considerations can also be applied to the
\textit{hydraulic conductivity} tensor, ${\sf K}_{ij}$, which is a 2nd rank
tensor that appears in \textit{Darcy's law}, for the general case of an
anisotropic porous medium, say,
\begin{equation}
           q_i=-{\sf K}_{ij} \PDX[j,h],
 \label{darcy}
\end{equation}
where $q_i$ denotes the $i$th component of the flux vector (= discharge per unit
area of porous medium per unit time), $h$ denotes the piezometric head, and the
vector $- {\Grad h}$ is the driving force. Here also, the ${\sf K}_{ij}$--matrix is
positive definite and symmetric, i.e., ${\sf K}_{ij}={\sf K}_{ji}$. It is
interesting to note that we may have cases with components ${\sf K}_{ij}\leq 0$ for
$i\neq j$. This is a consequence of the positiveness of the principal minor,
  \be
  K_{11}K_{22}-K_{12}^2 \geq 0, \quad \Rightarrow  \quad -\sqrt{K_{11}K_{22}}\leq
  K_{12} \leq \sqrt{K_{11}K_{22}}.
 \label{range}
 \ee

For an isotropic porous medium, ${\sf K}_{ij} = {\sf K} \delta_{ij}$, with
$\delta_{ij}$ denoting the \textit{Kronecker delta}.

 The above conclusions are
valid also for the coefficient of thermal conductivity, the coefficient of (mass)
diffusivity, and the dispersion coefficient of any extensive quantity transported in
the fluid that occupies the void space, or part of it.  For example:\vskip 3pt

\begin{enumerate}
\item \textbf{Mass Transport of a dissolved solute}. Such mass can be transported by
advection with the moving fluid (not considered here), by dispersion (as described
by (\ref{flux-c})) and by \textbf{(molecular) diffusion}. For a dilute system, the
latter mass flux is described by Fick's law (for a porous medium),
 \be
 J^{\gamma}_i=- {\cal D}^{\gamma\ast}_{ij} \PDX[j,c],
 \label{flux-cd}
  \ee
in which $ J_i^{\gamma}$ denotes the $i$th component of mass flux of a dissolved
$\gamma$-species (mass of $\gamma$ per unit area of porous medium per unit time),
$c$ denotes the concentration of that species, and ${\cal D}^{\gamma\ast}_{ij}$
denotes the coefficient of molecular diffusivity of $\gamma$ in the porous medium.
It takes into account the coefficient of molecular diffusion of $\gamma$ in the
fluid as a continuum, the porosity of the porous medium and the tortuosity of the
void space within the latter (e.g. Bear and Bachmat, 1990, p.\ 193).  The matrix
${\cal D}^{\gamma \ast}_{ij}$ is symmetric and positive definite.

\item \textbf{Heat transport}. Although heat can be transported also by the solid
matrix, we focus on the case in which the latter is thermally an insulator. Heat can
then be transported by advection with the moving fluid (not discussed here), by
thermal dispersion and by thermal conduction (which is a diffusive-type flux),
described by (the averaged) Fourier law,
\begin{equation}
           J^{\SCR{H}}_i=-\lambda^{\SCR{H} \ast}_{ij} \PDX[j,T],
 \label{fourier}
\end{equation}
in which ${ J}_i^{\SCR{H}}$ denotes the $i$th component of the heat flux by
conduction (heat per unit area of porous medium per unit time), $T$ is the
temperature, and $\lambda^{\SCR{H}\ast}_{ij}$ denotes the $ij$-component of the
coefficient of thermal conductivity of the porous medium. Again, the latter takes
into account the thermal conductivity of the fluid, as well as the porosity and
tortuosity of the void space occupied by the fluid. The matrix
$\lambda^{\SCR{H}\ast}_{ij}$ is symmetric and positive definite. \vskip 6 pt
 When the fluid occupying the void space is moving, heat is also
transported by
 advection, with the average velocity of the fluid. and by
\textit{thermal dispersion}. In analogy to (\ref{flux-c}), the \textit{thermal
dispersive flux} is expressed by
  \be
 J^{\SCR{H} \ast}_{disp, i}=- {\sf D}^{\SCR{H}\ast}_{ij} \PDX[j,T],
 \label{flux-cdisp}
  \ee
where $J^{\SCR{H}}_{disp, i}$ denotes the $i$th component of the thermal dispersive
flux vector (= heat per unit area of porous medium per unit time), and the vector $-
{\Grad T}$ is the driving force. The  ${\sf D}^{\SCR{H}\ast}_{ ij}$-matrix is
symmetric and positive definite.
\end{enumerate}

Although, we have extended the conclusion about the coefficients being positive
definite and symmetric, to a number of transport coefficients, there is a basic
difference between the coefficients of of hydraulic conductivity,   ${\sf K}_{ ij}$,
of mass diffusivity, ${\cal D}^{\gamma\ast}_{ij}$, and of thermal diffusivity
\textit{in a porous medium}, $\lambda^{\SCR{H} \ast}_{ij}$, as compared to those of
(mass) dispersion, ${\sf D}_{ij}$, and thermal dispersion, ${\sf D}^{
\SCR{H}}_{ij}$:
 \bitem
\item ${\sf K}_{ij}$ depends on the geometry of the void space through which
the water flows (tortuosity and width of pathways), in addition to fluid
properties,
such as density and dynamic viscosity.
\item ${\cal D}^{\gamma\ast}_{ij}$ and $\lambda^{\SCR{H} \ast}_{ij}$ depend on
the geometry of the fluid pathways, as well as the diffusivity and thermal
conductivity in the fluid.

In all these cases, the geometry of the fluid-occupied domain is expressed by a
scalar that represents the width of the pathways, and a 2nd rank tensor called
tortuosity that represents the effect of the tortuous fluid pathways. Bear
(1972, p. 111) showed that the tortuosity is a 2nd rank symmetric tensor.

\item The coefficients of mass and thermal dispersion, ${\sf D}_{ij}$ and
${\sf D}^{\SCR{H}\ast}_{ij}$, respectively,  are  functions not only of the geometry
of the void space, but also of the velocity field within the porous medium domain.
\eitem

We have mentioned several times that a considered fluid  may occupy only part
of the void space, at some fluid saturation (= volume of fluid per unit volume
of void space). Wherever a coefficient depends on the configuration of the void
space, it also depends on the saturation of the considered fluid. This aspect
is discussed in Sec.\ \ref{multi}.

The last two coefficients are discussed in the next section.

\section{The dispersivity, $a_{ijkl}$}\label{sec:2}

The dispersivity, $a_{ijkl}$, is related to the coefficient of dispersion ${\sf
D}_{ij}$ by (e.g. Bear, 1972, p.\ 610)
 \be
{\sf D}_{ij}=a_{ijkl} \frac{V_k V_l}{V}, \label{dispersivity}
 \ee
where $V_k$ denotes the $k$th component of the fluid's velocity vector, ${\V V}$,
with $V \equiv \vert {\V V} \vert$. The coefficient $a_{ijkl}$ is a fourth rank
tensor having the following properties:
\begin{enumerate}
 \item From the expression for the \textit{rate of entropy production}, $\dot{\cal S}$,
and following the discussion leading to (\ref{sigma-1}), we have
  \be
 \dot{\cal S} =Y_i X_i=\chi \left( - {\sf D}_{ij}\PDX[i,c]\right)
 \times \chi  \left(- \PDX[j,c]\right)
    =\chi^2 a_{ijkl}\PDX[i,c]\PDX[j,c] \frac{V_k V_l}{V} \geq 0,
    \label{sigma-2}
       \ee
from which it follows that $a_{ijkl}$ is positive definite. Thus, all
\textit{principal minors} of $a_{ijkl}$ are positive.
\item It has 3$^4 =81$ components in a 3-dimensional porous medium domain,
constrained by $2^{(3^2)}-1=511$ inequalities.
\item It is invariant under the following permutation of indices (see (\ref{sigma-2})),
 \be
  a_{ijkl} =a_{ijlk}, \quad a_{ijkl} = a_{jikl}.
 \label{aijkl-sym}
 \ee
Hence, only 36 components are \textit{independent} of each other. Furthermore, there
are $2^6-1=63$ constraining inequalities.
\end{enumerate}
Let us apply the above conclusions to an isotropic porous medium domain and to two
anisotropic ones.
\subsection{Isotropic porous medium}\label{isotropic}

For this case, the 36 independent components reduce to \textit{two}. This can be
shown  (Sirotine and Chaskolskaya, 1984, p.\ 651-2) by noting that the fourth rank
tensor satisfies the relationships  (\ref{aijkl-sym}) and is invariant under the the
action of full rotational (orthogonal) symmetry group O(3).
 In the case considered here, the $a_{ijkl}$-tensor can be represented by the
 matrix $p_{\alpha \beta}$, with
\begin{eqnarray}
\alpha,\beta= 1,2,3 \quad  \mbox{representing} \quad  xx,yy,zz,\quad\quad
 \alpha,\beta= 4,5,6 \quad \mbox{representing} \quad yz,zx,xy,\label{xx-yy}
 \end{eqnarray}
 is:
 \be
\left( \begin{array}{cccccc}
p_{11} & p_{12} & p_{12} & 0 & 0 & 0 \\
p_{12} & p_{11} & p_{12} & 0 & 0 & 0 \\
p_{12} & p_{12} & p_{11} & 0 & 0 & 0 \\
0 & 0 & 0 & p_{44} & 0 & 0 \\
0 & 0 & 0 & 0 & p_{44} & 0 \\
0 & 0 & 0 & 0 & 0 & p_{44}
\end{array} \right),\qquad p_{44}=\frac{p_{11}-p_{12}}{2}.
\label{matrix-is} \ee Denoting $p_{44}= (a_{\SCR{L}}-a_{\SCR{T}})/2$,
$p_{11}=a_{\SCR{L}}$, and $p_{12}= a_{\SCR{T}}$, we obtain:
 \be
 a_{ijkl} = a_{\SCR{T}} \delta_{ij}\delta_{kl} + \frac{a_{\SCR{L}} - a_{\SCR{T}}}{2}
           \left(\delta_{ik}\delta_{jl} + \delta_{il}\delta_{jk} \right)
\label{aijkl-d}
 \ee
and
\be
 {\sf D}_{im} = a_{\SCR{T}} V \delta_{im}+ \left(a_{\SCR{L}}-a_{\SCR{T}}\right)
           \frac{V_iV_m}{V}.
\label{aijkl-V}
 \ee
where the two independent coefficients are the \textit{longitudinal dispersivity},
$a_{\SCR{L}}$, and the \textit{transversal dispersivity}, $a_{\SCR{T}}$ (e.g. Bear,
1972, p.\ 611). Furthermore, from the positive definiteness of $a_{ijkl}$, it
follows
  \be
  a_{\SCR{L}} \geq 0, \quad a_{\SCR{T}} \geq 0. \label{def}
    \ee
\subsection{Axially symmetric porous medium}\label{anisotropic-4}

This case, also called \textit{axially symmetric anisotropy}, occurs, for example,
when the porous medium is made of many, relatively thin isotropic porous layers. The
same kind of material is obtained when we fill space by parallelepiped solid bodies,
say boxes, $a \times a \times b$, in the $x,y,z$, directions, respectively, with
equal spacing between all boxes.  The $z$-axis is an axis of symmetry. In this case,
the 36 independent components of $a_{ijkl}$ reduce to \textit{six} (Sirotine and
Chaskolskaya, 1984, p.\ 652),
 \be \left( \begin{array}{cccccc}
p_{11} & p_{12} & p_{13} & 0 & 0 & 0 \\
p_{12} & p_{11} & p_{13} & 0 & 0 & 0 \\
p_{31} & p_{31} & p_{33} & 0 & 0 & 0 \\
0 & 0 & 0 & p_{44} & 0 & 0 \\
0 & 0 & 0 & 0 & p_{44} & 0 \\
0 & 0 & 0 & 0 & 0 & p_{66}
\end{array} \right),\qquad p_{66}=\frac{p_{11}-p_{12}}{2},
\label{matrix-anis-ax} \ee where we have made use of (\ref{xx-yy}), taking the
$z$-axis as the axis of symmetry. The number 6 is reached by considering fourth rank
tensors that satisfy the relationships (\ref{aijkl-sym}) and are invariant under the
the action of the uniaxial symmetry group, $D_{\infty h}$.

Furthermore, because of the positiveness of the minors of the above matrix, the 6
constraints satisfy:
 \be p_{11}, p_{33}, p_{44} \geq 0, \quad  p_{11} \geq
p_{12},\quad p_{11} p_{33}
\geq p_{13}p_{31},\quad \det\left( \begin{array}{ccc}p_{11} & p_{12} & p_{13}\\
p_{12} & p_{11} & p_{13} \\
p_{31} & p_{31} & p_{33}\end{array} \right)\geq 0,  \label{xx-yy-zz}
\ee
where we have made use of (\ref{xx-yy}).

Another form of presenting $a_{ijkl}$, with the vector ${\V e}$ (components $e_i$)
denoting the direction of the axis of symmetry, is
\begin{eqnarray}
a_{ijkl} & = & a_1\delta_{ij} \delta_{kl}
       +\frac{a_2}{2}\{\delta_{ik}\delta_{jl}+\delta_{il}\delta_{jk} \}+
a_3 e_i e_j\delta_{kl}+a_4 e_k e_l\delta_{ij}+\nonumber \\
&&\frac{a_5}{2}\{e_i e_k\delta_{jl}+e_j e_k\delta_{il}+e_i e_l\delta_{jk}+
e_j e_l\delta_{ik}\}+a_6  e_i  e_j  e_k  e_l\;.\label{pp-oo}
\end{eqnarray}
Poreh (1965), using Robertson (1940), derived an expression for $a_{ijkl}$ which
missed 3 of the terms appearing in  (\ref{pp-oo}). The corresponding ${\sf
D}_{ij}$ is
\begin{eqnarray}
D_{ij}V &=& a_1\delta_{ij}(V_k V_k)+a_2V_iV_j+
a_3e_ie_j(V_k V_k)+a_4 (V_k e_k)(V_l e_l)\delta_{ij}+\nonumber \\
&&\frac{a_5}{2}\{e_i(V_k e_k) V_j + e_j(V_k e_k)V_i+e_i(V_l e_l)V_j+e_j
(V_le_l)V_i\}+a_6 e_i e_j(V_k e_k)(V_l e_l)\nonumber \\
             &=&
 e_i e_j\{a_6(V_ke_k)^2+a_3V^2\}+\delta_{ij}\{a_4(V_ke_k)^2+a_1V^2\}
                        + a_2 V_i V_j+\nonumber\\
&& + a_5(V_k e_k)\{e_i V_j + e_j V_i \}.\label{tt-yy}
\end{eqnarray}

As an example, consider the case of a layered porous medium domain. This is an
axially symmetric case, with the $z$-axis as axis of symmetry. Consider now uniform
flow \textit{normal} to the layers. This means: $e_1=e_2=0,e_3=1$ and
$V_1=V_2=0,V_3=V$. For this case, (\ref{tt-yy}) can be written as:
 \be
{\sf D}_{ij} =\left(\begin{array}{ccc} {\sf a}_{\SCR{THV}} & 0 & 0  \\
                    0 & {\sf a}_{\SCR{THV}} & 0 \\
                0&0&{\sf a}_{\SCR{LVV}}\end{array}\right) V, \quad
             \begin{array}{ll}
         {\sf a}_{\SCR{THV}}= &a_1+a_4,\\
    {\sf a}_{\SCR{LVV}}=&a_1+a_2+a_3+a_4+2a_5+a_6\end{array}\label{tt-yy-2}
 \ee
For uniform flow parallel to the layers, say, in the $+x$-axis, $e_1=e_2=0,e_3=1$
and $V_1=V, V_2=V_3=0$, (\ref{tt-yy}) can be rewritten as:
 \be
{\sf D}_{ij} =\left( \begin{array}{ccc} {\sf a}_{\SCR{LHH}} & 0 & 0  \\
         0 & {\sf a}_{\SCR{THH}} & 0 \\
      0 & 0 & {\sf a}_{\SCR{TVH}}\end{array}\right) V, \quad
            \begin{array}{ll}
         {\sf a}_{\SCR{LHH}}= &  a_1+a_2,\\
      {\sf a}_{\SCR{THH}}= &  a_1,\\
    {\sf a}_{\SCR{TVH}} = &a_1+ a_3.\end{array}
 \label{tt-yy-1}
 \ee

 Thus, to fully describe dispersion in a layered porous medium, when the flow is
 uniform and normal to the layers, we need one longitudinal and one transversal
dispersivities. For uniform flow \textit{in} the layers, we need (another) one
longitudinal and two transversal dispersivities.  In addition to these 5
coefficients, we need information on the direction of the axis of symmetry, i.e., we
need \textit{six} independent coefficients to completely describe the dispersion in
such a domain, when the flows are uniform, wither along the axis of symmetry or
normal to it. In the case of general flow through such domain, as shown earlier, we
need \textit{six} independent coefficients, plus information about the axis of
symmetry. Note that because of the axial symmetry, only one angle is required to
completely define the axis of symmetry.

It is interesting to note that Batchelor (1946), in his work on axisymmetric
turbulence, using a method based on invariance, introduced by Robertson (1940),
derived a general expression for ${\sf D}_{ij}$ that is based on \textit{five}
tensorial terms, $e_i e_j$, $\delta_{ij}$, $V_i V_j$, $V_i e_j$,  $V_j e_i$, and
four scalar functions, $C_i$:
 \be
{\sf D}_{ij}V=C_1({\V V}, {\V e}) e_i e_j + C_2({\V V}, {\V e})\delta_{ij}
   +C_3({\V V}, {\V e})V_i V_j + C_4({\V V}, {\V e})\{ V_i e_j+V_j e_i \}.
    \label{bat-2-9}
\ee
To construct the generic forms of the four scalar function $C_i$, where
$C_i=C_i({\V V}, {\V e}), i=1,..4$, such that they survive under the action of
axisymmetric group, noting that there are only two quadratic-in-$V$ scalar
invariants: $V^2$ and $(V_k e_k)^2$, we have:
\begin{eqnarray}
C_1= a_6(V_ke_k)^2+a_3V^2 ,\quad C_2=a_4(V_ke_k)^2+a_1V^2,\quad C_3=a_2,\quad
C_4= a_5 (V_k e_k). \label{bat--11}
 \end{eqnarray}
We have chosen the use of the same notation as in (\ref{tt-yy}) in order to
emphasize the explicit equivalence between the two expressions for ${\sf D}_{
ij}$. Poreh (1965) followed the same analysis as that presented by Batchelor
(1946). Altogether, the above comments explain why in the literature one
can find the
numbers four, five and six proposed by different authors. As we have shown, there is
no contradiction; ${\sf D}_{ij}$ is expressed by \textit{five} tensorial terms,
associated with \textit{four} quadratic-in-$V$ functions. The latter are constructed
by \textit{six} independent coefficients.

Bear and Bachmat (1990, p.\  215) also used Robertson's (1940) method and
reached the results of 2 independent coefficients for the isotropic case, and 6
for the axially symmetric one. However, they, erroneously, also assumed symmetry
permutations of the couples of indices $i,k$ and $j,l$, which led to only 5
independent scalar coefficients for the latter case. The reason for the
additional assumption of symmetry followed from the assumption of analogy
between dispersion and elasticity.
\subsection{Analogy with elasticity}\label{elasticity}

It is often stated that the relationship (\ref{dispersivity}) is analogous to
the relationship between the stress ($\sigma_{ij}$) and the strain
($\varepsilon_{ij}$) for an elastic solid:
 \be
 \varepsilon_{ij}=   C_{ijkl}\, \sigma_{kl},
\label{hook}
 \ee
in which $C_{ijkl}$ denotes components of the solid's elastic modulus, which is
a 4th rank tensor. Here, following the discussion leading to (\ref{sigma-1}),
the rate of entropy production, is expressed by:
 \be
\dot{\cal S}\equiv (\chi \sigma_{ij})( \chi \varepsilon_{ij}) =\chi^2 \sigma _{kl}
\left(C_{ijkl} \sigma_{ij} \right), \label{hook-2}
 \ee
which leads to the symmetry $C_{ijkl}=C_{klij}$,  in addition to the validity of
(\ref{aijkl-sym}).  Thus, for an isotropic solid, have
 \be
 C_{ijkl}=A_1 (\delta_{ik}\delta_{jl} + \delta_{il}\delta_{jk}) +
 A_2\delta _{ij}\delta_{kl},
 \label{2285}
 \ee
in which $A_1$ and $A_2$ are scalar coefficients, which are related to the
Lam\'{e} constants for the elastic solid. Thus, for an isotropic solid, because
these two coefficients are properties of the solid material only, they can be
expressed in terms of the two coefficients: Young's modulus of elasticity and
Poisson's ratio.

In the axially symmetric case (symmetry group $D_{\infty h}$), the total number (=
21) of independent components,  $C_{ijkl}$ is reduced to 5. However, in the case of
the dispersivity tensor, $a_{ijkl}$, we have shown that the number of independent
coefficients is 6 and not 5. Thus, the assumption of analogy is erroneous.

\subsection{Anisotropic porous medium with tetragonal symmetry}
\label{anisotropic-g}

We consider the case of an anisotropic porous medium with \textit{tetragonal
symmetry}. This kind of symmetry is referred to as D$_{4 h}$ (Sirotine and
Chaskolskaya, 1984, p.\ 651). As an example of such porous medium material, we may
consider one that is made up of orderly packed solid boxes $a\times a \times c$ with
equal spacing between the boxes in all directions (or cubes with 3 different
spaces). For this case, the dispersivity tensor can be described by the matrix
representation
 \be \left( \begin{array}{cccccc}
p_{11} & p_{12} & p_{13} & 0 & 0 & 0 \\
p_{12} & p_{11} & p_{13} & 0 & 0 & 0 \\
p_{31} & p_{31} & p_{33} & 0 & 0 & 0 \\
0 & 0 & 0 & p_{44} & 0 & 0 \\
0 & 0 & 0 & 0 & p_{55} & 0 \\
0 & 0 & 0 & 0 & 0 & p_{55}
\end{array} \right),
\label{matrix-anis} \ee where we have made use of (\ref{xx-yy}).

 In this case, the 36 independent components can be expressed by 7 independent
 parameters. Furthermore, because of the positiveness of the minors of the above matrix,
 we have the 10 constraints:
\[ p_{11}, p_{33}, p_{44}, p_{55} \geq 0, \quad  p_{11}^2 \geq p_{12}^2, \quad
p_{11} p_{33} \geq p_{13}p_{31},\quad\det \left( \begin{array}{ccc}
p_{11} & p_{12} & p_{13} \\p_{12} & p_{11} & p_{13} \\
p_{31} & p_{31} & p_{33}  \\\end{array} \right)\geq 0.
\]

This is an obvious extension of (\ref{tt-yy-1}), which presents the
case of flow parallel to the layers in an axially-symmetrical case.
 Altogether, we
need information on
 the directions in which the boxes, $a \times a \times c$, are
positioned in space
 (and this requires information on 2 angles), and then 3 coefficients:
(one
 longitudinal dispersivity coefficient and two different transversal
ones) for
 uniform flow in the $a$-direction, and 2 coefficients (one (different)
longitudinal dispersivity and
 one (different) transversal dispersivity) for uniform flow in the
$c$-direction Altogether, 7 independent coefficients
 (or 5, if the spatial directions are known).

Another (non-matrix) form of the dispersion tensor, ${\sf D}_{ij}$, can be
obtained
by considering  the invariants that can be constructed of the four vectors: the
velocity, ${\V V}$ and the three mutually orthogonal unit vector, ${\B \alpha}$,
${\B \beta}$, ${\B \gamma}$, such that
\[ \alpha_k \beta_k = \beta_k \gamma_k = \gamma_k \alpha_k=0,\quad
\alpha_k \alpha_k =\beta_k \beta_k =\gamma_k \gamma_k = 1,\]
 \be {\B \gamma}= {\B
\alpha} \times {\B \beta}, \quad {\B \beta}= {\B \gamma} \times {\B \alpha},\quad
{\B \alpha}= {\B \beta} \times {\B \gamma}.\label{zz-xx-cc}
 \ee
We also require that the symmetric tensor ${\sf D}_{ij} ( ={\sf D}_{ji})$
survives under the action of tetragonal group,
 \be
{\sf D}_{ij} ({\B \alpha})= {\sf D}_{ij} (-{\B \alpha}), \quad
{\sf D}_{ij} ({\B \beta})= {\sf D}_{ij} (-{\B \beta}), \quad {\sf D}_{ij}
({\B \gamma})= {\sf D}_{ij} (-{\B \gamma}), \quad {\sf D}_{ij} ({\B \alpha,
\beta, \gamma})= {\sf D}_{ij} ({\B \beta,\alpha, \gamma}). \label{kk-ll}
 \ee
With the above in mind, we start by listing all symmetric 2nd rank tensors built of
the  four vectors: ${\V V}$ and ${\B \alpha}$, ${\B \beta}$, ${\B \gamma}$:
 \be
\delta_{ij},\quad \alpha_i \alpha_j, \quad\beta_i \beta_j,\quad \gamma_i \gamma_j,
 \quad V_i V_j, \label{pp}
 \ee
  \be \alpha_i V_j + \alpha_j V_i, \quad \beta_i V_j
+ \beta_j V_i, \quad \gamma_i V_j + \gamma_j V_i. \label{oo}
 \ee
  \be \alpha_i
\beta_j + \alpha_j \beta_i, \quad \beta_i \gamma_j + \beta_j \gamma_i, \quad
\gamma_i \alpha_j + \gamma_j \alpha_i.\label{ll} \ee
Actually, because of the relationship (\ref{zz-xx-cc}), only two of these vectors
suffice to completely define  ${\sf D}_{ij}$, provided, these vectors are chosen
consistently. Here, once we select ${\B \alpha}, {\B \beta}$, the vector ${\B
\gamma}$ is well defined. Accordingly, we obtain
 \be
\delta_{ij},\quad \alpha_i \alpha_j, \quad\beta_i \beta_j,
 \quad V_i V_j, \label{ppp}
 \ee
  \be \alpha_i V_j + \alpha_j V_i, \quad \beta_i V_j
+ \beta_j V_i, \quad \alpha_i \beta_j + \alpha_j \beta_i. \label{ooo} \ee

Next we have to select coefficients such that when multiplying each of the above
terms, contributes a quadratic in $V$ term to the expression of ${\sf D}_{ij}$.
Furthermore, the resulting ${\sf D}_{ij}$ has to satisfy (\ref{kk-ll}). With
this last requirement, the terms in ({\ref{ll}) have no contribution. The first
three terms in (\ref{ppp}), require coefficients of the form
\[
A_1 \left( (V_k \alpha_k)^2 +  (V_k \beta_k)^2 \right) + A_2 V^2 ,
\]
in which the $A_i$'s take on different values for each of the first three terms in
(\ref{ppp}). For  the last term in (\ref{ppp}), the coefficient  is a constant. As
for the last term in (\ref{ooo}), it can be shown that only \textit{one} scalar
invariant can serve as a coefficient, as otherwise, condition (\ref{kk-ll}) will be
violated. Thus, we obtain:
\[ B\left( (\alpha_k V_k)(\alpha_i V_j + \alpha_j V_i) +
(\beta_k V_k)(\beta_i V_j + \beta_j V_i)\right).\] As to the last term in
(\ref{ooo}), its contribution is:
\[ C (\alpha_k V_k)(\beta_k V_k)(\alpha_i \beta_j + \alpha_j \beta_i).\]

 Altogether, we arrive at the final  expression, in terms of 7 coefficients:
 \begin{eqnarray}
  V D_{ij} &=&  \left( A_1 \left( (\alpha_k V_k)^2 +  (\beta_k V_k)^2)\right) + A_2 V^2
 \right) \delta_{ij} \nonumber \\
&& +  \left( A_3 \left( (\alpha_k V_k)^2 +  (\beta_k V_k)^2)\right) + A_4 V^2
 \right) \left(\alpha_i \alpha_j  + \beta_i \beta_j \right)\nonumber \\
&& + A_5\left( (\alpha_k V_k)(\alpha_i V_j + \alpha_j V_i) +
(\beta_k V_k)(\beta_i V_j + \beta_j V_i)\right) \nonumber \\
 && + A_6(\alpha_k V_k)(\beta_k V_k)(\alpha_i \beta_j + \alpha_j \beta_i) + A_7 V_i V_j .
\label{Dij-an}
\end{eqnarray}

As an example, consider the case of a porous medium made of boxes, $a \times a
\times c$,
 with $a$ in the $x$ and $y$ directions and $c$ in the $z$-direction,  and ${\B
\alpha} = (1,0,0), {\B \beta}= (0,1,0), {\B \gamma}=(0,0,1)$. For uniform flow in
the $x$ or $y$ directions, i.e., $V_1=V, V_2=V_3=0$, (\ref{Dij-an}) can be written
as:
 \be
D_{ij} = \left( \begin{array}{ccc} a_{\SCR{LHH}} & 0 & 0  \\
                                   0 & a_{\SCR{THH}} & 0 \\
                                    0 & 0 & a_{\SCR{TVH}} \end{array} \right) V,\quad
                                    \begin{array}{ll}
                                    a_{\SCR{LHH}}= & A_1+A_2+A_3+A_4+2A_5+A_7,\\
                                    a_{\SCR{THH}}= & A_1+A_2+A_3+A_4,\\
                                    a_{\SCR{TVH}}= & A_1+A_2.
                                    \end{array}
 \label{tt-yy-3}
 \ee
For uniform flow  in the $+z$-axis,  $V_1=V_2=0, V_3=V$, (\ref{Dij-an}) can be
rewritten as:
 \be
D_{ij} =\left( \begin{array}{ccc} a_{\SCR{THV}} & 0 & 0  \\
                                   0 & a_{\SCR{THV}} & 0 \\
                                   0 & 0 & a_{\SCR{LVV}} \end{array} \right) V,\quad
                                    \begin{array}{ll}
                                    a_{\SCR{THV}}= & A_2 + A_4,\\
                                     a_{\SCR{LVV}} = & A_2 +A_7.\end{array}
 \label{tt-yy-1}
 \ee

 Thus, to fully describe dispersion in a layered porous medium, when uniform flow
 takes place
 in the $+x$ or $+y$ directions, we need one longitudinal and two transversal
dispersivities--in the horizontal direction and in the vertical one. For uniform
flow \textit{in} the vertical ($+z$-)direction, we need (another) one longitudinal
and one transversal dispersivities. Altogether we need 5 coefficients, plus
information---two angles---on the directions of the axes, i.e., we need
\textit{seven} independent coefficients to completely describe the dispersion (i.e.,
${\sf D}_{ij}$) in such a porous medium.

Needless to add that in all cases, the actual values of the various coefficients
have to be determined \textit{experimentally}. The knowledge about the number of
independent coefficients in each case facilitates the design of the experiments.

The analysis presented above for the cases of isotropic  and anisotropic porous
media, can be extended to any other types of anisotropy.

 \section{Coupled processes}
 \label{coupled}

\subsection{Conjugate fluxes and forces}\label{congugate}
The generalized Newton's law that describes molecular flux of linear momentum,
Fourier's law, that describes conductive heat flux, and Fick's law that
describes diffusive mass flux of a solute (= molecular diffusion), are
particular cases of the general linear law for phenomena of transport in a fluid
phase (e.g. Bear and Bachmat, 1990, p.\ 99)
\begin{equation}
J_i^n = - \sum_{j=1}^{3} L_{ij}^{nn}{\PDX[j,\Phi^{n}]}, \quad i,j = 1, 2, 3,
\quad (\mbox{or}\quad =x,y,z),  \label{<2.2.92>}
\end{equation}
where $J_i^n$ denotes the $i$th component of the flux of an extensive quantity
$E^n$ of a phase, $\Phi^n$ is a state variable associated with $E^{n}$ and $L_{
ij}^{nn}$ is a coefficient of proportionality.

The three linear diffusive flux laws mentioned above, state that a nonuniform
distribution of the state variable, $\Phi^n$ (e.g., temperature), produces a flux of
only the corresponding extensive quantity (e.g., heat). However, {\em experimental
evidence} suggests that, in principle, gradients of state variables, $\Phi^r$,
$r\neq n$, corresponding to other extensive quantities, may also contribute to the
flux of $E^n$. Such phenomena are referred to as {\em coupled phenomena}. An example
of such phenomenon is the {\em Soret} (or {\em thermodiffusion}) {\em effect}, in
which mass flux of a solute in a liquid phase is produced by a temperature gradient,
in addition to the flux produced by the gradient of the solute's concentration
according to Fick's law. Another example is the {\em Dufour effect}, in which heat
flux is caused by a concentration gradient, in addition to the heat flux caused by
temperature gradient, according to Fourier's law. Actually, the diffusive flux of a
component, as expressed by Fick's law, manifests interdependence between the
concentrations of components in a multicomponent system,
 \be
J^\beta_i= -\sum_{\alpha=1}^{N}  \sum_{\j=1}^{3} {\cal D}_{ij}^{\alpha\beta}
\PDX[j,c^\beta], \quad \alpha, = 1,2,,\ldots, N.
 \label{Fick-multi}
  \ee
where $N$ denotes the number of dissolved chemical species, and $c^\beta$
denotes the concentration of the $\beta$-species. Using the terminology
introduced above, we have to regard the mass of a dissolved $\beta$-species as
one of the $E^n$ extensive quantities.

All the above diffusive flux expressions are valid within the fluid that
occupies the void space of a porous medium, or part of it, i.e., at the
\textit{microscopic level} of description. By appropriate averaging, we obtain
the corresponding flux laws at the {\em macroscopic}, or {\em averaged} level.
In fact, they have the same linear form as (\ref{<2.2.92>}), except that $J_i^n$
denotes the $i$th component of the flux of an extensive quantity $E^n$ of a
phase, \textit{per unit area of porous medium}, $\Phi^n$ is the macroscopic
state variable associated with $E^{n}$, and the coefficient $L_{ij}^{nn}$ takes
into account the porosity and the tortuosity of the porous medium. For short,
we shall leave the same symbols also for the macroscopic interpretation, as
our discussion focuses on forces, fluxes, and coefficients at the macroscopic
level.

We consider $\ell$ extensive quantities, and $\ell$ corresponding  fluxes
produced by the {\em thermodynamic forces},
\begin{equation}
X_j^r = -\PDX[j,\Phi^r], \quad
 r=1,2,\ldots,\ell, \quad
\label{<2.2.96>}
\end{equation}
Each diffusive flux $Y_i^r$ and  the corresponding force $X_i^r$ are chosen such
that they are \textit{thermodynamically conjugate force and flux} (Sec.\
\ref{sec:1}). For the set of coupled fluxes, we have:
\begin{equation}
Y_i^q = \sum_{r=1}^{\ell} \sum_{j=1}^{3} L_{ij}^{qr} X_{j}^{r}, \quad i = 1,
2, 3; \quad q = 1, 2, \ldots, \ell. \label{<2.2.98>}
\end{equation}

From the {\em phenomenological equations} (\ref{<2.2.98>}), it follows that the flux
of an extensive quantity, $E^q$, is a single-valued function of {\em all} the
(coupled) thermodynamic forces associated with $E^q$.  They express linear
relationships between fluxes and thermodynamic forces.

Equation (\ref{<2.2.98>}) can be extended to coupled phenomena of a higher
order, i.e., when the transport is described by coefficients that are tensors
of 3rd and higher ranks. As examples, we may mention the coupling between
piezoelectric and piezomagnetic phenomena discussed by Fel (2002).

Before continuing with the discussion of coupled processes, let us make the
following important comment. Although the hydraulic conductivity $K_{ij}$,
appearing in Darcy's law, and all the coefficients $L_{ij}^{qr}$ appearing in
the equation that describes coupled diffusive fluxes, are 2nd rank tensor
coefficients, there is a basic difference between them. Although Darcy (1856)
obtained his law as an empirical one,  it is actually a simplified version of
the averaged momentum balance equation of a fluid. This law exists only at the
macroscopic (porous medium) level, describing advective flux in a porous medium
domain, while all other diffusive flux laws exist in a fluid continuum, i.e.
at the microscopic level. When averaged, they become macroscopic diffusive flux
laws (of heat, mass of a solute, etc.) in a porous medium domain. Coupling as
discussed above exists only among diffusive-type fluxes.

In 1851, Stokes postulated that in (\ref{<2.2.92>}), the coefficients $L_{ij}^
{qq}$ are symmetric with respect to the permutation of the coordinates $i$
and $j$, i.e.
\begin{equation}
L_{ij}^{qq} = L_{ji}^{qq}. \label{<2.2.99>}
\end{equation}

This postulate states that the transformation of a unit force along one axis
into a conjugate flux along another axis remains \index{Conjugate flux}
unaltered when those axes are interchanged. In Sec.\ \ref{sec:1}, we have
discussed the three coefficients ${\sf D}_{ij}$, ${\sf K}_{ij}$ and ${\sf D}^{
\SCR{H}\ast}_{ij}$.

Bear and Bachmat (1990, p.\ 226) have extended the concept of coupled processes
to the macroscopic level in a porous medium domain. They suggested that the
concept of coupling, remains valid also at the macroscopic level, provided the
coefficients involved, $L^{qr}_{ij}$ corresponding to a phase continuum, be
multiplied by the tortuosity of the porous medium.

Let us focus on the {\em cross coefficients} for $q \neq r$, which give the flux
of $E^q$ caused by the force, ${\V X}^r$, associated with the gradient of $e^r$
($\equiv$ the density of $E^r$). Employing the {\em principle of microscopic
reversibility of processes}, and methods of statistical mechanics, Onsager
(1931) showed that for the linear equations (\ref{<2.2.98>}), {\em and provided
a proper choice is made for the fluxes}, ${\V Y}^q$, {\em and forces}, ${\V
X}^r$, {\em the phenomenological coefficients are also symmetric in} $r$
{\em and} $q$, i.e.
\begin{equation}
L_{ij}^{qr} = L_{ji}^{rq} , \qquad q \neq r. \label{<2.2.100>}
\end{equation}
This follows from the symmetry of the composite matrix.  For the example, the
composite matrix for $\ell=2$ takes the form:

\be  \left( \begin{array}{cc}
 \left( L^{11}_{ij} \right)  & \left( L^{12}_{ij} \right)  \\
 & \\
\left( L^{21}_{ij} \right) & \left( L^{22}_{ij} \right)   \\
\end{array} \right).\ee
These, relationships are known as\, {\em Onsager's}, or {\em Onsager--Casimir's,
reciprocal relations}.
 They express a relationship between any pair
of cross-phenomena (e.g., thermal diffusion and Dufour effect) arising from
simultaneously occurring irreversible processes \index{Dufour
effect}\index{Irreversible processes} (e.g., heat conduction and molecular
diffusion).  According to Onsager, the reciprocal relations, (\ref{<2.2.99>}),
hold under two conditions (e.g., Bear and Bachmat, 1990, p.\ 103):
\begin{enumerate}
\item The relationship between each individual flux and its
conjugate thermodynamic force is \textit{linear}.
\item The fluxes, ${\V Y}^q$, and their conjugate
\index{Conjugate force}\index{Conjugate flux} forces, ${\V X}^q$, should be
selected such that
\begin{equation}
\dot{{\cal S}} = \sum_{q=1}^{\ell} \sum_{i=1}^{3} Y_i^q X_i^q, \label{<2.2.102>}
\end{equation}
where $\dot{{\cal S}}$ is the rate of entropy production of the system
(De Groot, 1963, Ch.\ 1).
\end{enumerate}
Equation (\ref{<2.2.102>}) implies that each ${\V Y}^q$ and its conjugate force,
${\V X}^q$, must be of the same tensorial rank (not necessarily a vector, as
indicated by the single subscript).

By the {\em 2nd law of thermodynamics}
\index{Second law of thermodynamics}
\begin{equation}
\dot{{\cal S}} \geq 0\quad \rightarrow\quad
\sum_{(q,r,i,j)} L_{ij}^{qr} X_j^r X_i^q \geq 0. \label{<2.2.104>}
\end{equation}

Hence, a necessary condition for the validity of (\ref{<2.2.104>}) is that all
diagonal matrices $ L_{ij}^{qq}$ are positive definite. Let us consider
three special cases of porous media.

\subsection{Isotropic porous medium, $O(3)$}

 For an isotropic medium,  using the example of $i,j,=1,2,3$ and $\ell=2$, e.g.,
coupled fluxes of heat and solute, we can write the coefficients $L^{qr}_{ij}$
in the matrix form,
 \be \left(
         \begin{array}{cc}
 \left( \begin{array}{ccc} a & 0 & 0 \\
                           0 & a & 0 \\
                            0& 0& a   \end{array} \right)  &
 \left( \begin{array}{ccc} c & 0 & 0 \\
                           0 & c & 0 \\
                            0& 0& c   \end{array} \right)  \\
                          &                        \\
 \left( \begin{array}{ccc}c & 0 & 0 \\
                         0 & c & 0 \\
                          0 & 0 & c  \end{array} \right)  &
 \left( \begin{array}{ccc} d & 0 & 0 \\
                           0 & d & 0 \\
                            0& 0& d   \end{array} \right)
\end{array}\right) \ee
with $a,c,d$ denoting coefficients, e.g., $a\equiv L^{11}_{11}$, and $d\equiv
L^{22}_{11}$. We have here 3 independent coefficients, with the inequalities
$a,d\geq 0$, and $ad>c^2$.

\subsection{Anisotropy---with uniaxial symmetry, $D_{\infty h}$ }
For this case, again, with $i,j,=1,2,3$ and $\ell=2$, the matrix of
coefficients, can be represented as
 \be
\left(\begin{array}{cc}
 \left(\begin{array}{ccc} a & 0 & 0 \\
                           0 & a & 0 \\
                            0& 0& d   \end{array} \right)  &
 \left(\begin{array}{ccc}  b & 0 & 0 \\
                          0 & b & 0 \\
                          0 & 0 & c  \end{array} \right)  \\
                          &                        \\
\left(\begin{array}{ccc}  b & 0 & 0 \\
                             0 & b & 0 \\
                             0 & 0 & c  \end{array} \right)   &
\left(\begin{array}{ccc}  e & 0 & 0 \\
                                   0 & e &0 \\
                                  0 & 0 & f \end{array} \right)
\end{array} \right)
\ee
Altogether, we have here 6 independent coefficients, which are governed by 6
inequalities:
\[
a,d,e,f \geq 0, \quad ae \geq b^2, \quad df \geq c^2.
\]
\subsection{Anisotropy--with orthorhombic symmetry, $D_{2h}$}
For the example of $\ell=3$, the matrix takes the form,
  \be
\left(\begin{array}{ccc}
 \left(\begin{array}{ccc} a & 0 & 0 \\
                           0 & b & 0 \\
                           0& 0&  c  \end{array} \right)  &
 \left(\begin{array}{ccc} m & 0 & 0 \\
                          0 & n & 0 \\
                          0 & 0 & s  \end{array} \right)  &
 \left(\begin{array}{ccc} r & 0 &0 \\
                           0 & p& 0 \\
                            0& 0 & q   \end{array} \right)  \\
                          &                        \\
 \left(\begin{array}{ccc} m & 0 & 0 \\
                           0 & n & 0 \\
                            0& 0& s   \end{array} \right)  &
 \left(\begin{array}{ccc}d & 0 & 0 \\
                          0 & e & 0 \\
                          0 & 0 & f  \end{array} \right)  &
 \left(\begin{array}{ccc} u & 0 & 0 \\
                           0 & v & 0 \\
                            0& 0& w   \end{array} \right)  \\
                          &                        \\
  \left(\begin{array}{ccc} r & 0 & 0 \\
                           0 & p & 0 \\
                            0& 0& q   \end{array} \right)  &
 \left(\begin{array}{ccc}u & 0 & 0 \\
                          0 & v & 0 \\
                          0 & 0 & w  \end{array} \right)  &
 \left( \begin{array}{ccc} g &0 & 0 \\
                           0 & h & 0 \\
                            0& 0& i   \end{array} \right)  \\
\end{array}\right) \ee
Here  we have 18 independent coefficients, satisfying the 18 inequalities:
\bea
a,b,c,d,e,f,g,h,i\geq 0,\nonumber
\eea
\bea
ad\geq m^2,\quad be\geq n^2,\quad cf\geq s^2, \quad ag \geq r^2, \quad bh\geq
p^2,\quad  ci\geq q^2, \quad dg\geq u^2, \quad eh\geq v^2, \quad fi\geq w^2.
\nonumber
\eea
For any $\ell$, the number of independent components is:
\begin{eqnarray}
\mbox{for symmetry group}\quad &O(3):&\quad\frac{1}{2}\ell (\ell+1).\nonumber\\
\mbox{for symmetry group}\quad &D_{\infty h}:&\quad\ell (\ell+1).\nonumber\\
\mbox{for symmetry group}\quad &D_{2h}:&\quad\frac{3}{2}\ell (\ell+1).\nonumber
\end{eqnarray}

It it is interesting to note that the material symmetries, $D_{4 h}$, and $D_{
6 h}$, with horizontal honeycomb cross-section, which are close $D_{\infty h}$
have also $\ell (\ell + 1)$ independent coefficients.

\vspace{-.3cm}
\section{Multiple fluid phases}\label{multi}
So far, the discussion has been related to phenomena of transport in
\textit{saturated flow}, i.e., in a fluid that fills up the entire pore space.
However, often, a number of fluid phases occupy the void space simultaneously,
each at at a certain saturation, where the saturation, $S_\alpha$ of an
$\alpha$-phase is defined as the ratio between the volume of the $\alpha$-phase
and the volume of the pore space, with $\sum_{\alpha} S_\alpha =1$. Under such
conditions, considering phenomena of transport within a fluid phase, neglecting
interphase transfers of the considered extensive quantities across (microscopic)
interphase boundaries, each of the components: $K_{ij}$, $a_{ijkl}$, and all
$L^{qr}_{ij}$'s, is a function of the phase saturation, $S_\alpha$. The reason
for this dependence stems from the fact that the geometric features of the phase
occupied portion of the pore space, e.g. the tortuosity, varies with the
saturation.

For example, for an anisotropic porous medium, with orthorhombic symmetry and
with $x,y,z$ principal directions, $k_{xx}=f_x(S_w)$, $k_{yy}=f_y(S_w)$,
$k_{zz}=f_z(S_w)$, but still, the $K_{xy}=f_{xy}(S_w) = K_{yx}(S_w)$, etc.
Another example is the dispersivity. $a_{ijkl}$. For an isotropic porous medium,
we have $a_{\SCR{L}}=f_{a\SCR{L}}(S_w)$, $a_{\SCR{L}}=f_{a\SCR{T}}(S_w)$, with
$f_{a\SCR{L}} \neq f_{a\SCR{T}}$.

\vspace{-.5cm}
\section{Summary} \label{summary}
We have studied a number of transport coefficients that appear in the laws that
govern advective, diffusive and dispersive fluxes of extensive quantities that
are transported within a phase occupying the void space of a porous medium
or part of it. We focused on an anisotropic porous medium. Making use of the
basic features of tensors of 2nd and 4th rank, we have determined for each
considered transport coefficient, the number of independent components. This
information will enable further research on the relationships among the various
dispersivity components. It will also facilitate the analysis of experiments
that are conducted in order to determine the numerical values of these
components by inverse methods.

In a separate article, Bear and Fel (2009) discuss the minimum number of experiments
that is required in order to determine the values of the six $a_i$ moduli for the
axially symmetric case.

\vspace{-.5cm}
\section{Acknowledgement}\label{ack}
The authors wish to thank Prof. Alexander H.-D. Cheng, Department of Civil
Engineering, University of Mississippi, Dr. Peter C.\ Lichtner, Los Alamos
National Laboratory, and Prof. M.\ Poreh, Department of Civil and Environmental
Engineering, Technion--Israel Institute of Technology, for fruitful discussions.
The research was partly supported by the Kamea Fellowship program.

\end{document}